\documentclass[letterpaper,twocolumn,english]{IEEEtran}
\usepackage[T1]{fontenc}
\usepackage{babel}
\usepackage{amsmath}
\usepackage{amssymb}
\usepackage{graphicx}
\usepackage[unicode=true,
 bookmarks=false,
 breaklinks=false,pdfborder={0 0 1},backref=section,colorlinks=false]
 {hyperref}
\makeatletter

\pdfpageheight\paperheight
\pdfpagewidth\paperwidth

\providecommand{\tabularnewline}{\\}

\usepackage{babel}

\makeatother
\begin{document}
\title{Inverse prediction of capacitor multiphysics dynamic parameters using
deep generative model}
\author{\IEEEauthorblockA{Kart-Leong Lim, Rahul Dutta, and Mihai Rotaru\\
Institute of Microelectronics, \\
\textit{Agency for Science, Technology and}\\
\textit{Research (A{*}STAR)}\\
Singapore, Singapore \\
 \{limkl, duttar, mihai\_dragos\_rotaru\}@ime.a-star.edu.sg }}
\maketitle
\begin{abstract}
Finite element simulations are run by package design engineers
to model design structures. The process is irreversible meaning every minute structural adjustment requires a fresh input parameter run. In this paper, the problem of modeling changing (small) design structures through varying input parameters is known as inverse prediction. We demonstrate inverse prediction on the electrostatics field of an air-filled capacitor dataset where the structural change is affected by a dynamic parameter to the boundary condition. Using recent AI such as deep generative model, we outperformed best baseline on inverse prediction both visually and in terms of quantitative measure. 
\end{abstract}


\section{Introduction}

Most multiphysics simulation \cite{hennigh2021nvidia,raissi2019physics,lim2022physics}
are centered around the forward input-output model and well studied
for decades. In electrostatics, we can model the relationship between
the Cartesian coordinates input and the electrostatics field output
of a capacitor using Laplace's equation \cite{9740189,nagel2011solving,lim2022physics}.
In fluid dynamics, we can model the spatial-temporal input to a heat
sink viscosity output using Burger's equation \cite{hennigh2021nvidia}.
However in a multi-billions industry such as in semiconductor, the
development of forward models are usually well established and made
available by commercial softwares. Computing multiphysics forward
models such as using finite element method (FEM) software requires
licensing. Also, it is impractical to run a new multiphysics design
for each dynamic parameter changes to the forward model. On the other
hand, performing inverse prediction/modeling \cite{9908139,lee2020dirty,Lebreux2010FastIP,schafer2022generalization,kim2021dpm}
is an ill-posed AI/ML problem that is very challenging to solve for
both the industry and researchers. In this work, we propose the use
of generative models such as the variational autoencoder (VAE) \cite{kingma2014stochastic}
and autoencoder (AE) \cite{hinton2006reducing} to model dynamic parameter
multiphysics simulation. Thus, we have a latent representation of
the multiphysics simulation that models infinte dynamic parameter
changes. We demonstrate in the experiment that inverse prediction
in the latent space will have reduced error compared to traditional
inverse prediction.

\section{Background}

The electrostatic field of an air-filled capacitor \cite{sadiku2015analytical}
can be described using collocation points and boundary points on a
2D grid in Fig 1. The collocation points are described by Laplace
equation and the boundary points by boundary conditions . We consider
five boundary conditions $C1$ to $C5$ defined using parameters ($a,b,d,V_{o}$)
along a 2D axis ($x,y$) in Fig 1. Instead of traditionally fixing
all the boundary parameters, we allow dynamic parameter $d=\left[0,1\right]$
to vary, which changes the electrostatic field extensively as seen
in Fig 2. In this work, we require a set of training data which comprises
of electrostatic fields $V$ and dynamic parameter $d$ which can
be obtained by commercial FEM softwares (e.g. ANSYS or MATLAB) or
PDE solver (e.g SOR \cite{nagel2011solving} or PINN \cite{9740189,lim2022physics,raissi2019physics}).

\begin{figure*}
\begin{centering}
\includegraphics[scale=0.8]{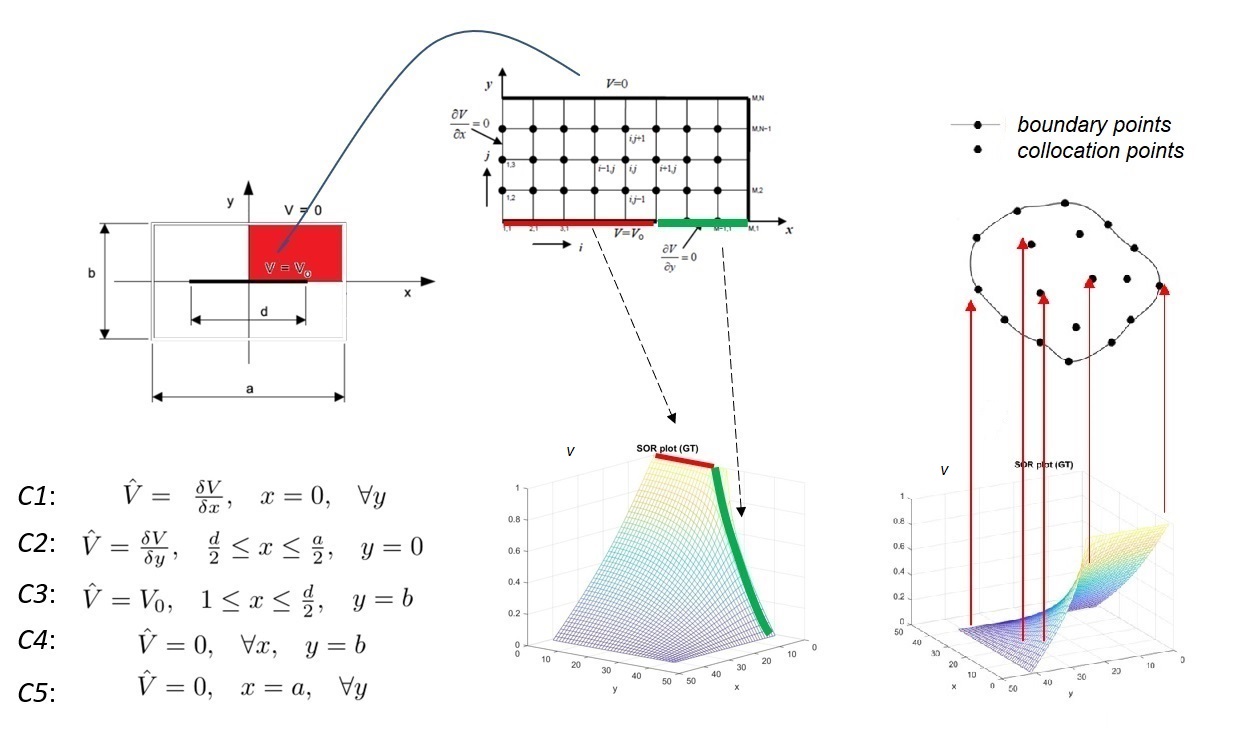} 
\par\end{centering}
\caption{Modeling the boundaries of an electrostatic field (an air-filled capacitor)
using five boundary conditions. In particular, $d$ affects boundary
conditions $C2$ and $C3$.}
\end{figure*}

\section{Problem statement: }

\subsection{Fullspace approach}

A trained regression model map electrostatics fields $V_{1...m}$
as input samples and their corresponding dynamic parameter $d\left(V_{1...m}\right)$
as output samples in eqn (1). Inverse prediction on the trained regression
model $\phi$ then predicts an unknown output $V_{i}$ when presented
with a given dynamic parameter input $d(V_{i})$. We can formulate
the objective of inverse prediction below which we refer to as the
fullspace approach:

\begin{equation}
\begin{array}{c}
\underset{\hat{V_{i}}}{\arg\min}\;\left[\hat{V_{i}}\phi-d(V_{i})\right]\\
\\
where\;\;V_{1...m}\phi=d\left(V_{1...m}\right)\\
\\
\end{array}
\end{equation}

The dimensionality of the input can reach a complexity as high as
$V_{i}\in\mathbb{R}^{I\times I}$ e.g. $I=401$, given the output
$d(V_{i})\in\mathbb{R}^{1}$. Thus, inverse prediction in eqn (1)
is an ill-posed problem due to three main problems: 
\begin{verse}
i) The root finding of $\hat{V_{i}}$ is difficult due to having too
many coefficients to minimize.

ii) A good initial estimate is often necessary.

iii) If training size $m$ is insufficient, it will lead to many redundant
coefficients for $\phi$. 
\end{verse}
\begin{figure*}
\includegraphics[scale=0.28]{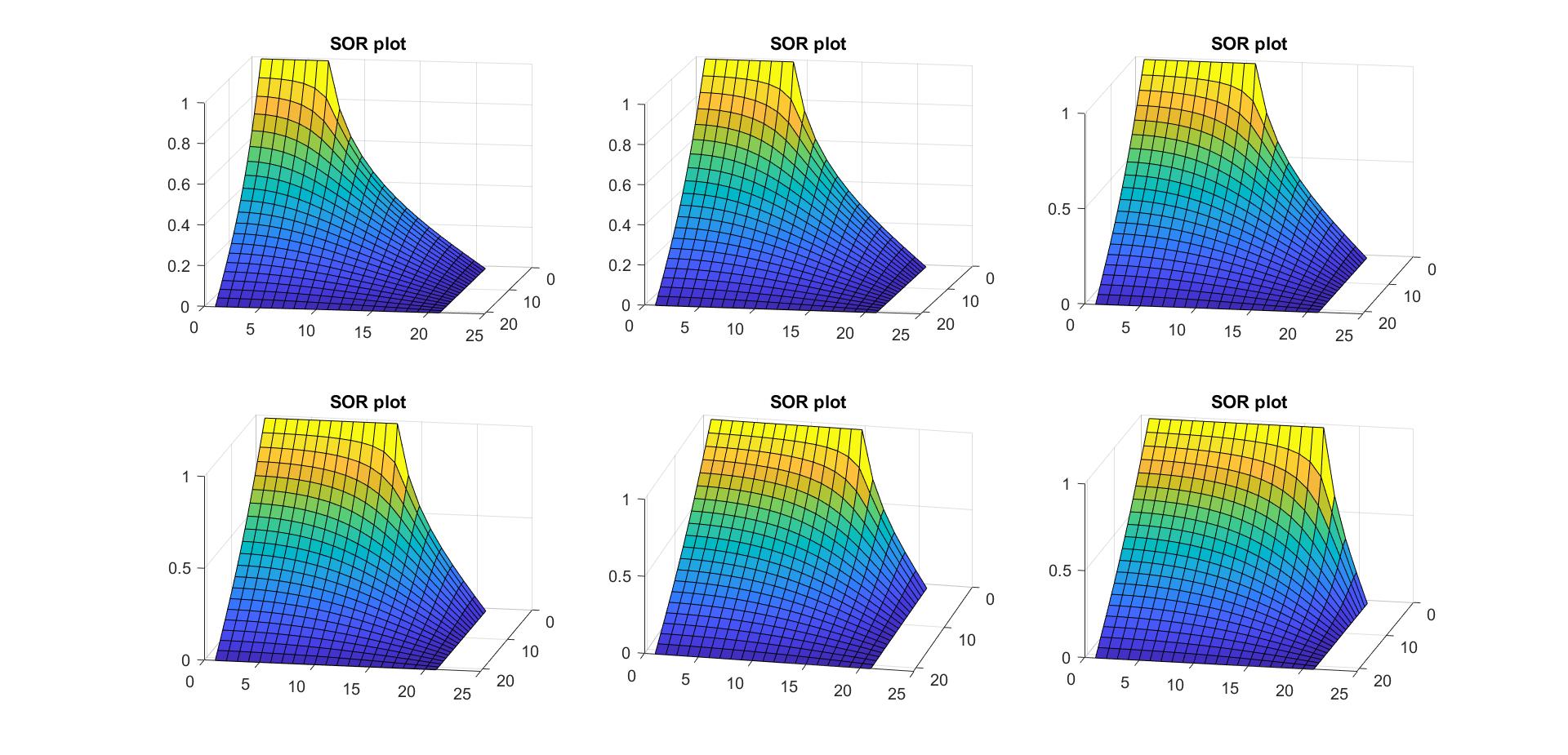}

\caption{Groundtruth $V$ plots of an air-filled capacitance, corresponding
to dynamic parameter $d=\left[0.3,0.4,...,0.8\right]$.}
\end{figure*}

\begin{figure*}
\begin{centering}
\includegraphics[scale=0.7]{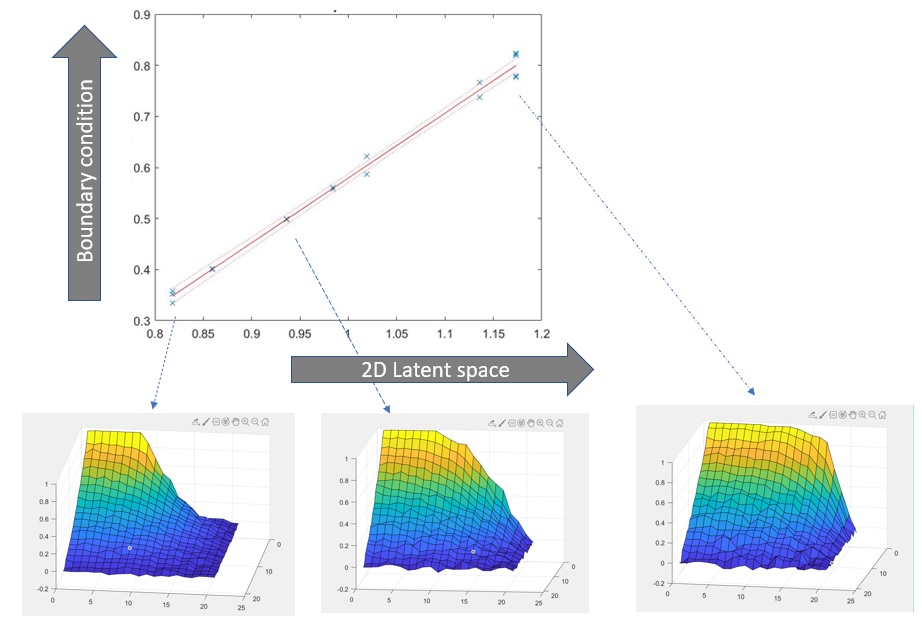} 
\par\end{centering}
\caption{$V$ plots modeled as points in the latent space. A regression model
$\phi$ is learnt on these points with their corresponding $d$. Then,
$\phi$ can be used to inverse predict an unknown latent point for
a new $d$.}
\end{figure*}

\section{Proposed method}

\subsection{Latent approach}

We proposed a latent approach as summarized in Fig 3 to address the
aforementioned problems. The main advantages of the proposed method
are: 
\begin{verse}
i) Lower error for modeling $V$: - using unsupervised deep learning
as feature extraction.

ii) Faster computational time: - both regression and inverse prediction
work less hard due to reduced dimensionality. 
\end{verse}
The latent approach uses eqn (2) for inverse prediction. Also, we
note that the dimension of $V_{1...m}\in\mathbb{R}^{I\times I}$ in
the latent space denoted as $L(V_{1...m})\in\mathbb{R}^{1\times Z}$
is much smaller i.e. $Z\ll I\times I$.

\begin{equation}
\begin{array}{c}
\underset{\hat{L(V_{i})}}{\arg\min}\;\left[\hat{L(V_{i})}\phi-d(V_{i})\right]\\
\\
where\;\;L(V_{1...m})\phi=d\left(V_{1...m}\right)\\
\\
\end{array}
\end{equation}

We detail our latent approach (Fig 4) as follows:

\textbf{Deep generative model (training)}: During training phase,
an unsupervised dataset of $V_{1...n}$ plots alone ($d$ is not used)
is made available and a VAE \cite{kingma2014stochastic,lim2020} or
AE \cite{hinton2006reducing,song2013auto} is trained. This is illustrated
by the blue solid arrow path.

\textbf{Regression model (training)}: A second supervised dataset
is presented whereby $L(V_{1...m})$ with known corresponding $d(V_{1...m})$
is used to train a regression model. It is shown using the blue dashed
arrow path connecting the green box $\phi$.

\textbf{Inverse prediction (testing)}: During testing, the red dashed
arrow path connecting the green box $\hat{V_{i}}$ recovers an unknown
$\hat{L(V_{i})}$ for a new $d(V_{i})$.

\textbf{Reconstruction (testing)}: Lastly, the test sample $L(V_{i})$
is fed to the decoder to reconstruct $V_{i}$ denoted as $V_{i}'$.

\begin{figure*}
\begin{centering}
\includegraphics[scale=0.77]{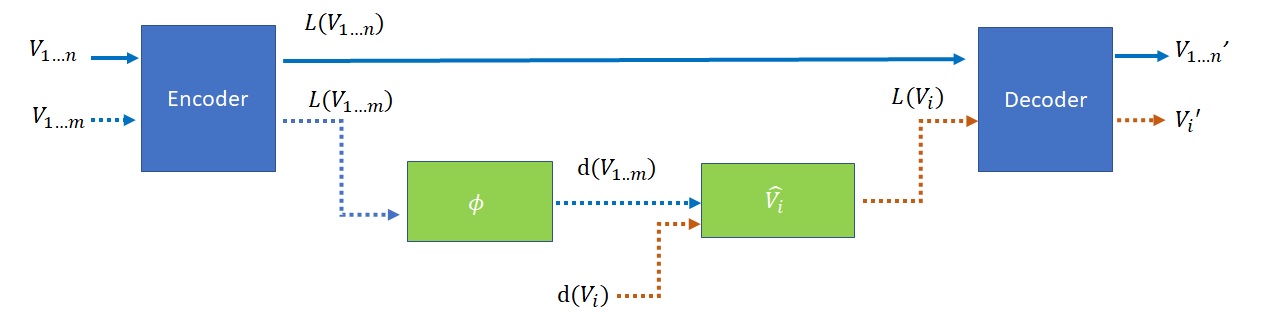} 
\par\end{centering}
\caption{Proposed latent approach for inverse prediction.}
\end{figure*}

\subsection{Deep generative models}

The latent space of AE denoted as $L^{AE}(V)$ is straightforward
defined using the hidden layer $\mu$ as shown in Fig 5. Whereas the
latent space of VAE denoted as $L^{VAE}(V)$ further consist of an
additional hidden layer $\sigma$ corrupted by Gaussian noise $\mathcal{N}(0,1)$.

\begin{equation}
\begin{array}{c}
L^{AE}(V)=\mu\\
\\
L^{VAE}(V)=\mu+\sigma\cdot\mathcal{N}(0,1)\\
\\
\end{array}
\end{equation}

While AE trains a \textbf{one-to-one} mapping of $V$ to $\mu$ (and
vice-versa), it is poor at reconstructing the targeted $V$ when $\mu$
is not longer accurate. The novelty of VAE comes from the fact that
it can train a \textbf{one-to-many} mapping of $V$ to $\mu$. Whereby
$\mu$ is learnt together with many draws of Gaussian in eqn (3).
This is a useful asset especially when the observed latent sample
is noisy in real cases.

VAE achieve this \textbf{one-to-many} mapping through KLD loss ($\mathcal{L}^{KLD}$)
on top of \textbf{one-to-one} mapping using reconstruction loss ($\mathit{\mathcal{L}^{REC}}$)
in eqn (4). 
\begin{equation}
\begin{array}{c}
\mathcal{L\mathrm{^{\mathit{REC}}}}=-\frac{1}{2}\left(V-V'\right)^{2}\\
\\
\mathcal{L}^{KLD}=-\frac{1}{2}\left(\sigma^{2}+\mu^{2}-\ln\sigma^{2}-1\right)\\
\\
\end{array}
\end{equation}

\begin{figure*}
\begin{centering}
\includegraphics[scale=0.55]{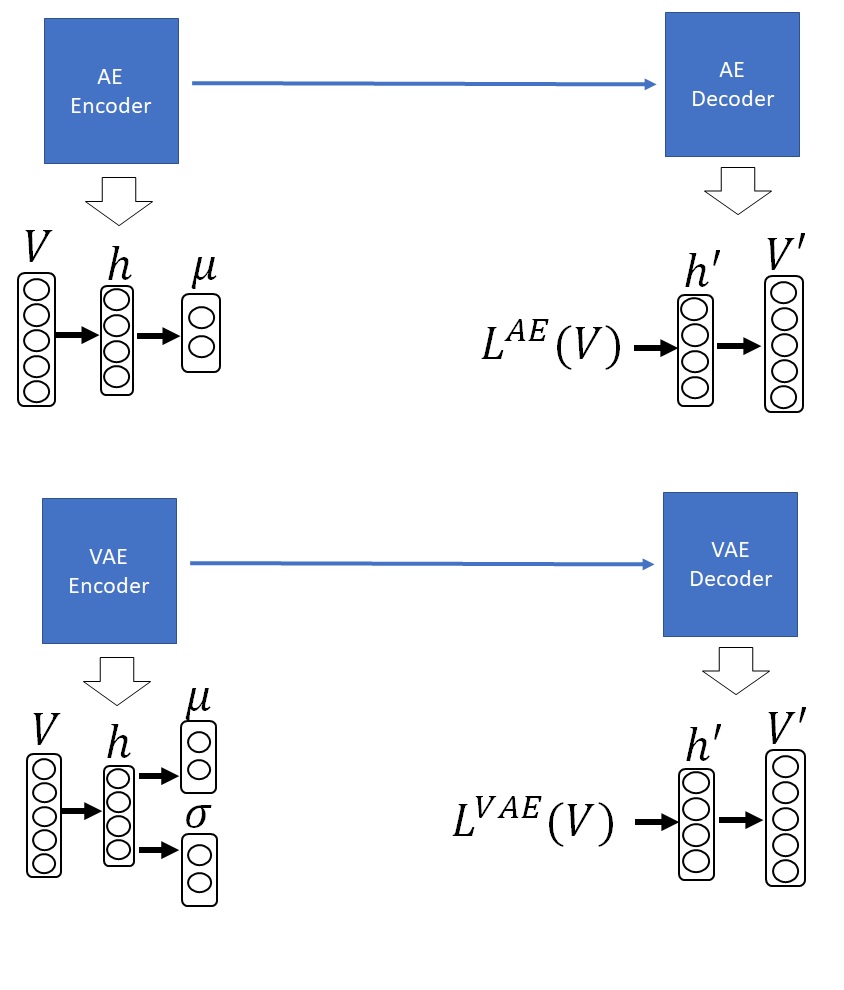} 
\par\end{centering}
\caption{Architecture of deep generative models: AE vs VAE.}
\end{figure*}

\section{Experiments}

We target the effectiveness of inverse prediction under noisy condition
on electrostatic field simulation. We propose using VAE in comparison
with strong and weak baselines. We refer to strong baseline as the
latent approach based on AE and weak baseline using fullspace approach.
We show that our proposed method consistently outperforms both baselines
under metric benchmark. We use the parameters in Table 1 when training
VAE and AE for producing the results in Fig 6-9.

\subsection{Noisy condition for initial estimate}

We refer to $X$ as the clean initial estimate for $V_{1...m}$ or
$L(V_{1...m})$ both corresponding to $d=0.5\pm0.2$. $G$ refers
to Gaussian noise with zero mean and variance, $e=[0.01,...1.0]$.
The noisy initial estimate $Y$ when corrupted by additive white Gaussian
noise (AWGN) is given as

\begin{equation}
\begin{array}{c}
Y=X+G\\
\\
G\sim\mathcal{N}(0,e)\\
\\
\end{array}
\end{equation}

Fig 6 mainly shows that a latent approach (AE, VAE) is more robust
than fullspace approach when using $Y$ instead of $X$. When $e$
is low, the fullspace approach shows error reproducing boundary condition
$C2$. When $e$ is large, the fullspace approach fails entirely to
produce an acceptable $V$ plot. Also, the output of AE becomes less
smooth than VAE as $e$ increases (which we shall make a detailed
comparison later on).

Visually, the plots of AE and VAE under noise are not smooth in Fig
6. To improve this issue, we use Adam to train the loss functions
for VAE. We compare VAE trained using Momentum \cite{pmlr-v28-sutskever13}
versus VAE trained using Adam \cite{kingma2014adam} in Fig 7 and
see that the latter produce significantly smoother plots very close
to groundtruth.

\begin{figure*}
\includegraphics[scale=0.3]{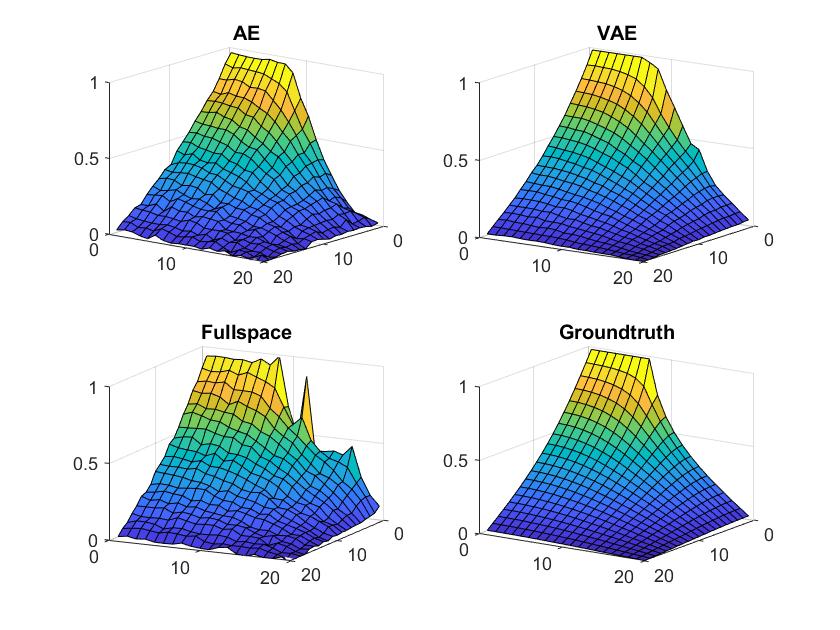}\includegraphics[scale=0.3]{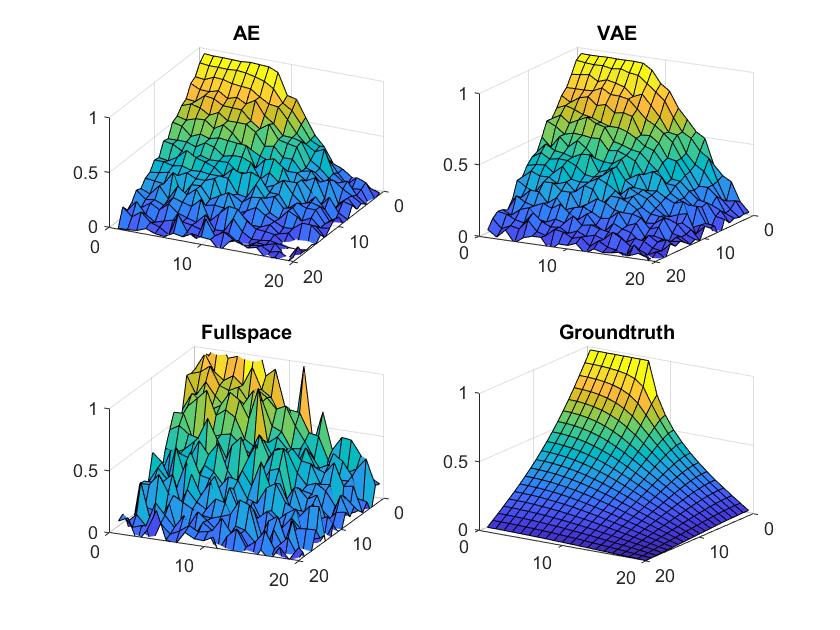}

\caption{Reconstruction of $V'$ plots ($d=0.36$) when presented with AWGN
($e=0.01$ and $e=0.1$) in eqn (5).}
\end{figure*}

\begin{figure*}
\includegraphics[scale=0.3]{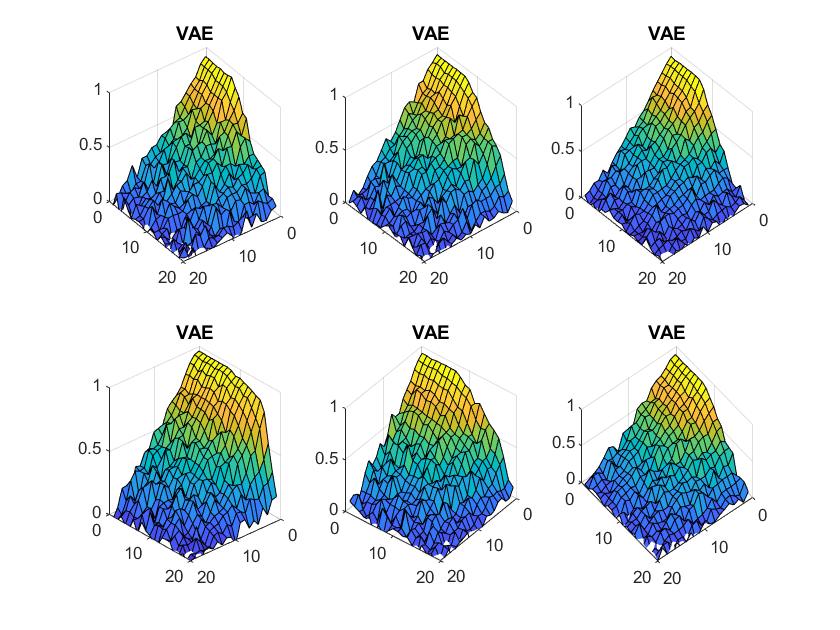}\includegraphics[scale=0.3]{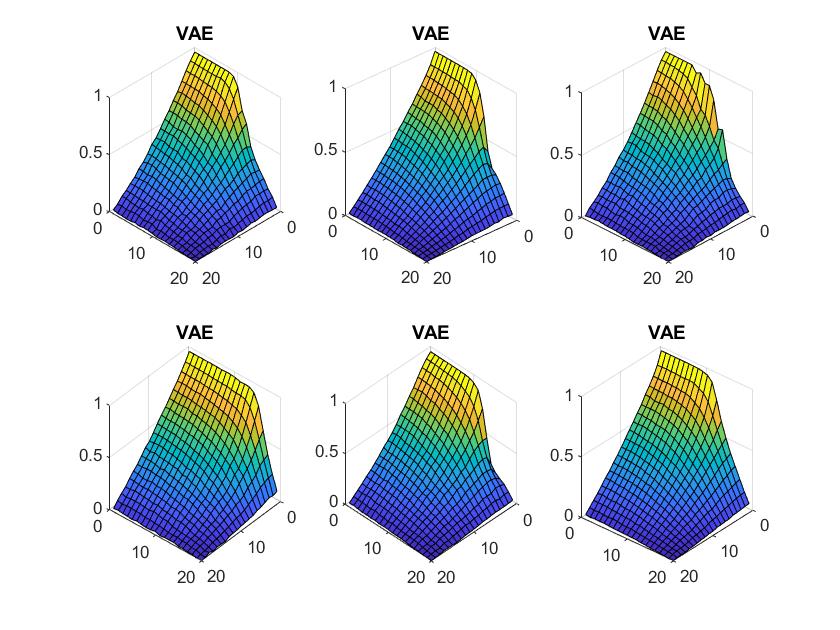}

\caption{Plots of $V$ for different $d$ with $e=0.1$. (Left plots) Momentum.
(Right plots) Adam.}
\end{figure*}

\subsection{Metric benchmark }

In our findings (based on Momentum for both AE and VAE), as the initial
estimate becomes noiser ($e$ increases from 0.01 to 0.2), the decoder
output of AE starts to become less smooth than VAE. In Fig 8, we use
a quantitative measure based on the sum-of-square-difference (ssd)
at each node of a 21x21 grid to capture this behavior.

\begin{equation}
ssd=\sum_{i=1}^{T}(V_{i}-\hat{V_{i}})^{2}
\end{equation}

Under low noise level of $e=0.01$, AE has much lower error than VAE.
However, as we increase the noise level from 0.1 towards 1.0, VAE
consistently produces less error than AE across all $d$.

Although not a direct comparison with AE, we see that when comparing
Fig 9 and Fig 8, VAE trained using Adam produce much lower error than
using Momentum for ssd across all noise level.

\begin{figure*}
\begin{centering}
\includegraphics[scale=0.22]{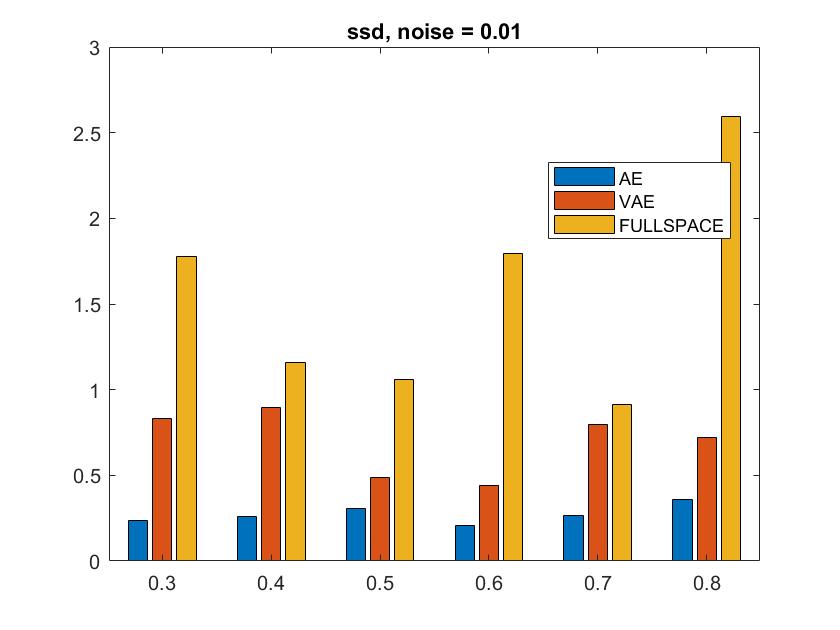}\includegraphics[scale=0.22]{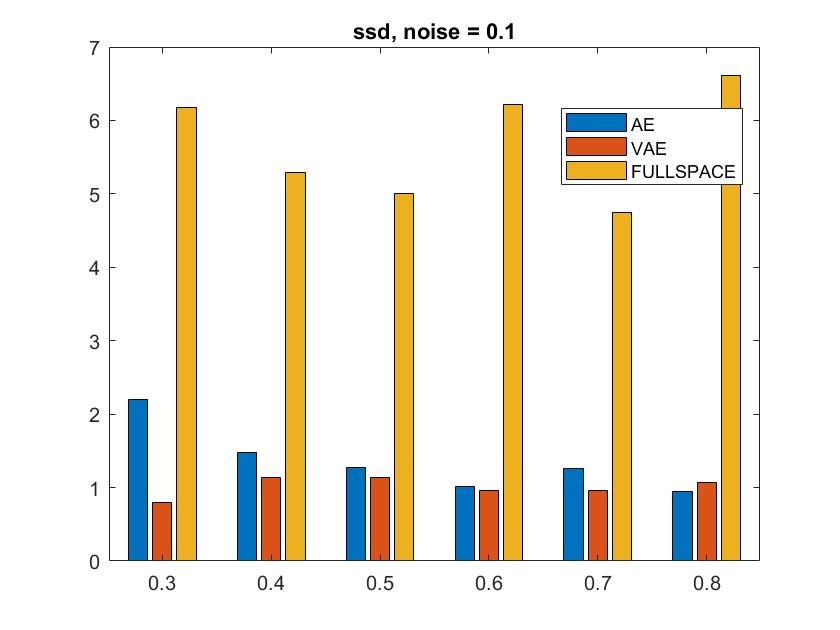} 
\par\end{centering}
\begin{centering}
\includegraphics[scale=0.22]{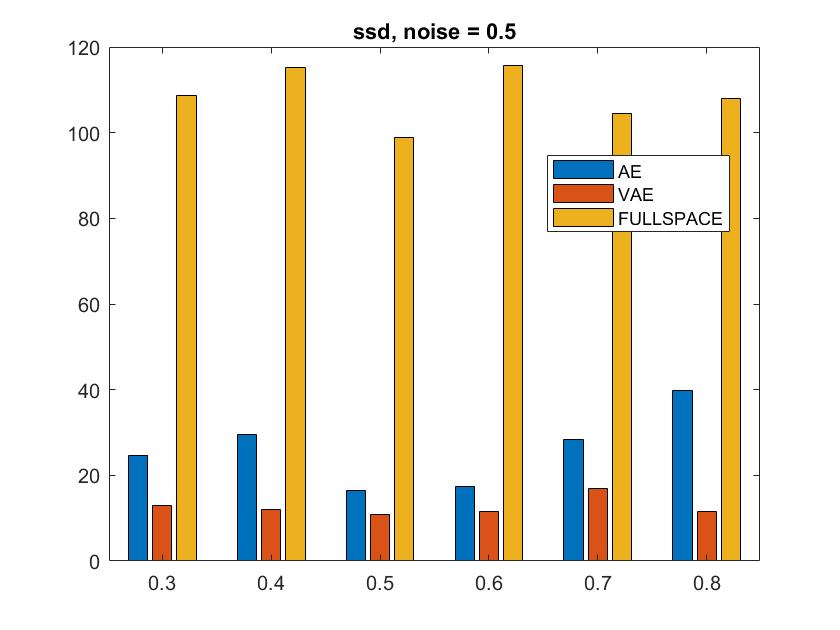}\includegraphics[scale=0.22]{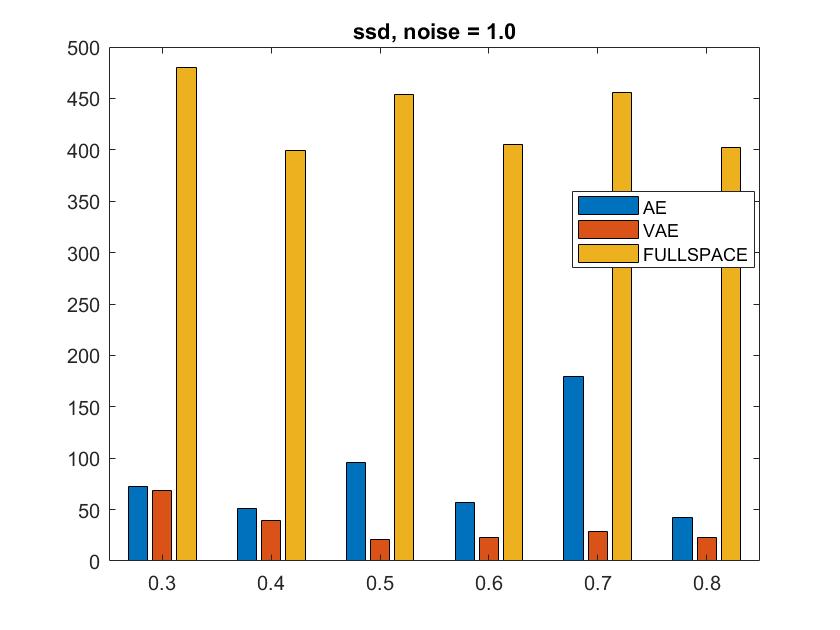} 
\par\end{centering}
\caption{Measuring sum-of-square-difference error at the decoder.}
\end{figure*}

\begin{figure*}
\begin{centering}
\includegraphics[scale=0.22]{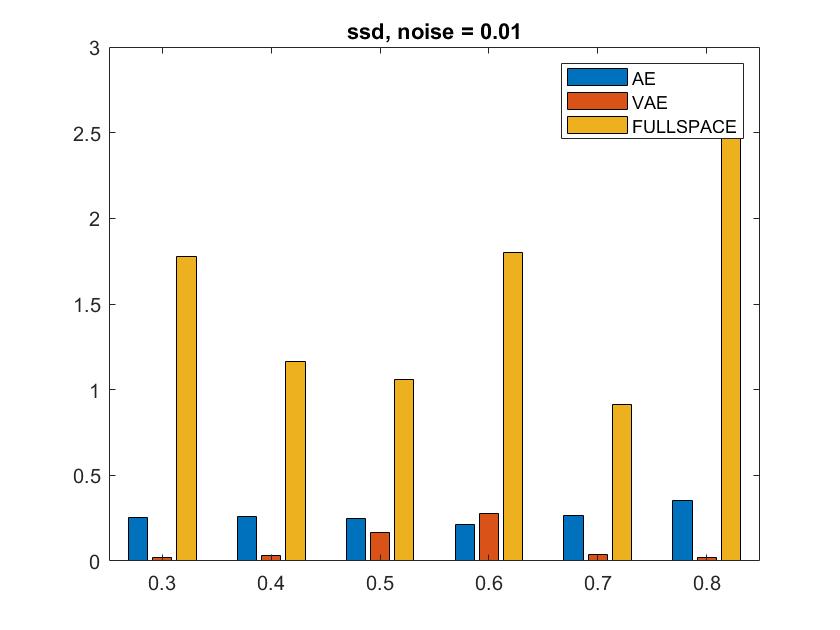}\includegraphics[scale=0.22]{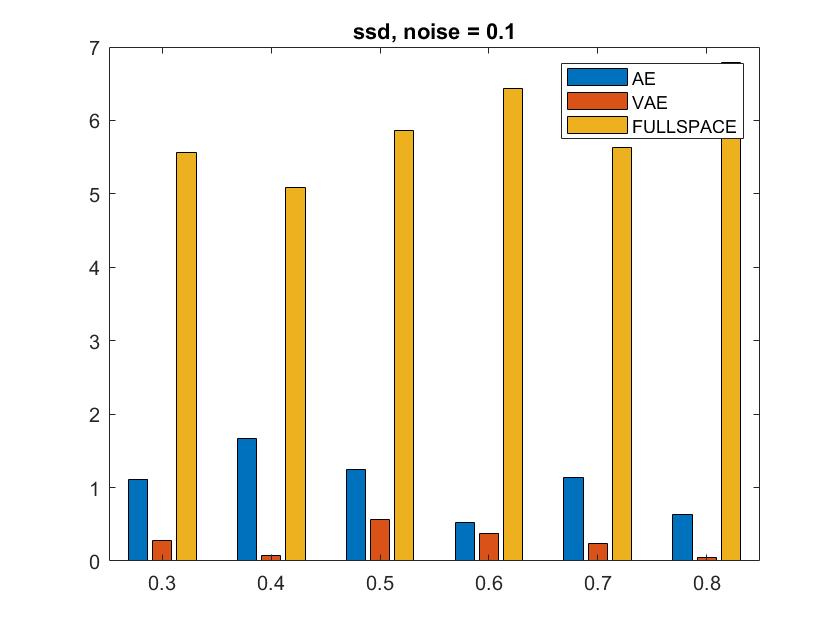} 
\par\end{centering}
\begin{centering}
\includegraphics[scale=0.22]{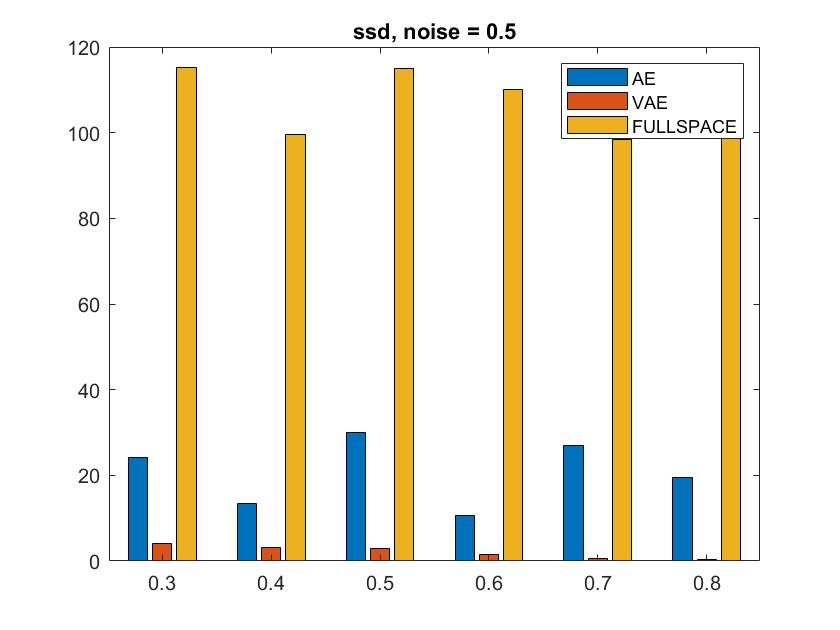}\includegraphics[scale=0.22]{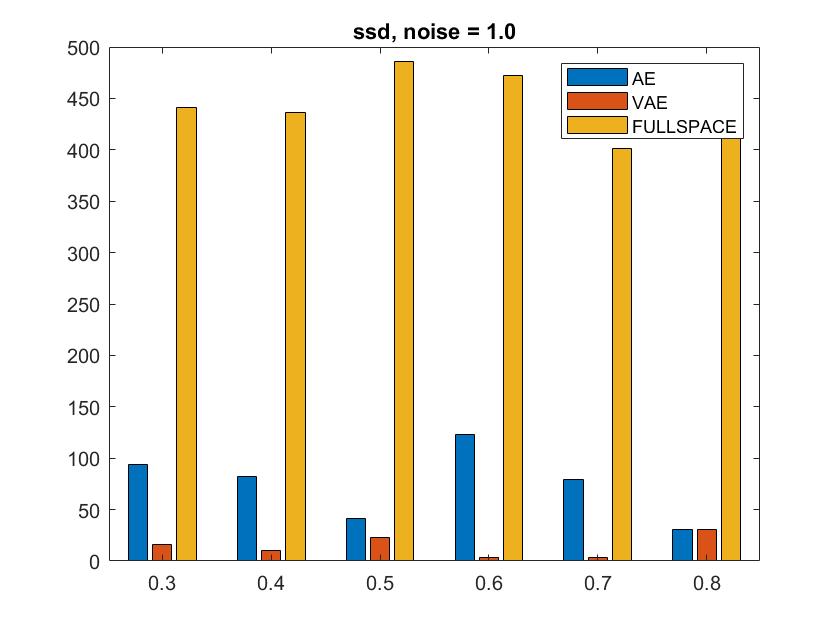} 
\par\end{centering}
\centering{}\caption{Measuring sum-of-square-difference error at the decoder (VAE trained
with Adam).}
\end{figure*}

\subsection{CPU Time}

We show in Table 2 the execution time taken for each operation in
the latent and fullspace approach. In particular, performing regression
and inverse prediction each took approximately 10x to 20x longer for
the fullspace approach. When we use the entire training and test data,
the fullspace approach takes 10x slower to compute than either AE
or VAE. This is due to both latent approach using a much smaller dimension
at 20 vs 441 when performing regression and inverse prediction.

\begin{table}
\caption{Parameters for deep generative model}
\begin{centering}
\begin{tabular}{|c|c|}
\hline 
 & Parameters\tabularnewline
\hline 
\hline 
Architecture  & 441-200-20-200-441\tabularnewline
\hline 
Activation function  & tanh\tabularnewline
\hline 
Max iterations  & 20,000\tabularnewline
\hline 
Gradient learner  & Momentum, Adam\tabularnewline
\hline 
Learning rate  & $1e^{-3}$, $1e^{-5}$\tabularnewline
\hline 
Minibatch size  & 20, 100\tabularnewline
\hline 
Train samples  & 120\tabularnewline
\hline 
Test samples  & 6\tabularnewline
\hline 
\end{tabular}
\par\end{centering}
\end{table}

\begin{table}
\caption{Computational time}
\begin{centering}
\begin{tabular}{|c|c|c|c|}
\hline 
CPU time (ms)  & Fullspace  & AE  & VAE\tabularnewline
\hline 
\hline 
encoder  & -  & 1  & 1\tabularnewline
\hline 
regression  & 1  & 0.2  & 0.5\tabularnewline
\hline 
inverse  & 180  & 4  & 5\tabularnewline
\hline 
decoder  & -  & 2  & 2\tabularnewline
\hline 
\hline 
total  & 1142  & 148  & 163\tabularnewline
\hline 
\end{tabular}
\par\end{centering}
\end{table}

\section{Conclusion}

We propose an inverse prediction on capacitor multiphysics in the
latent space. It allows the user to use a dynamic parameter to reconstruct
the exact electrostatic field distribution that accurately corresponds
to the boundary condition. Our proposed method is mainly based on
a deep generative model such as VAE. We showed that using sophisticated
gradient learner such as Adam our method can reconstruct field plots
very close to groundtruth.

 \bibliographystyle{IEEEtran}
\phantomsection\addcontentsline{toc}{section}{\refname}\bibliography{allmyref}

\end{document}